# Wavelength-Agnostic Metasurface Design for Next-Generation 2D Photodetectors

AYUSH M. JAMDAR,[1] RITURAJ,[2,*] SRINI KRISHNAMURTHY,[3,4] VIDYA PRAVEEN BHALLAMUDI,[3] SIVARAMA KRISHNAN,[3,**]

[1]*Department of Electrical Engineering, Indian Institute of Technology Madras, India*
[2]*Department of Electrical Engineering, Indian Institute of Technology Kanpur, India*
[3]*Department of Physics, Indian Institute of Technology Madras, India*
[4]*Sivananthan Laboratories, Bolingbrook, IL 60440, USA*
*[rituraj@iitk.ac.in](mailto:rituraj@iitk.ac.in)*
**[srkrishnan@iitm.ac.in](mailto:srkrishnan@iitm.ac.in)*

**Abstract:** This research presents a versatile method for designing two-dimensional photonic crystal-based dielectric metasurfaces. We optimize their geometric patterns to achieve specific functionalities like near-perfect absorption at chosen wavelengths. This technique can also be applied to tailor reflectance and transmittance for any optical mode and wavelength. Our method combines Covariance Matrix Adaptation (CMA) and Rigorous Coupled Wave Analysis (RCWA) for efficient design and optimization. We demonstrate its effectiveness by realizing simple, high-performance structures using conventional materials. This approach holds promise for applications in quantum technologies, optical communication, and nonlinear optics.

## 1. Introduction

In their seminal study [1], Piper *et al.* demonstrated that a photonic crystal structure (PCS) can significantly enhance the absorption of 1550 nm light in a graphene monolayer from a mere 2.3% [2] to a total absorption of 100% by inducing resonance. This is in stark contrast to the traditional photodetectors which often require a few-micron-thick absorber for efficient light absorption. However, this comes with the caveat that a thicker absorber leads to increased dark current lowering device sensitivity. By enabling efficient absorption in ultra-thin layers with PCS can lead to photodetectors with inherently ultra-low dark current as well as higher operating temperatures [3-5].

A key insight from this approach to device design [1] is that the optical performance depends solely on the geometric parameters of the metastructure, both the parameters that govern its long-range ordering, its periodicity, as well as those that define its unit cell - thickness, and pattern type - for fixed material composition. The dependence on material properties and their limitations are overcome and transcended by this powerful tool for designing resonances at any desired wavelengths to tailor optical responses beyond that achievable by conventional approaches, hitherto.  This strategy towards device design impacts the development of photonic devices for a wide range of applications with unique needs.

Total absorption of narrowband around a single wavelength is required in sensing and optical communications [6]. Ostensibly, in dual-wavelength scenarios such as non-degenerate time-resolved pump-probe type measurements [7], nonlinear harmonic conversion devices such as second harmonic generation (SHG) requiring intense primary input light beams which are often converted to relatively weak harmonic outputs at desired wavelengths [8], or essential quantum optic photon sources, bi- or entangled-photon, sources which rely on energy-momentum conserving nonlinear optical parametric down conversion where a pump photon following its interaction with material is converted to a pair of useful daughter photons at wavelengths different from the parent pump, double resonant designs are much needed, yet, rare to find. In optical communication systems, simple designs for seamless channel switching and efficient operation at both 1.3 µm and 1.55 µm wavelengths are much needed [9], as are add-drop optical filters that selectively transmit specific wavelengths while rejecting others [10]. Furthermore, tunable implementations of such devices exploiting unique material properties, such as those offered by phase change materials, PCMs, to retain the high-optical performance of metastructure designs are well-poised to disrupt hyper-spectral imaging [11]. Thus, the ability to design resonances through metasurfaces is crucial to meet both the unique and diverse requirements these evidently important applications place on tailoring device responses, leading to enhancements in their performance, often beyond the limits of primitive designs and architectures.

From a practioner's perspective, methods to design these metasurfaces can be classified into forward and inverse design. Traditionally, forward design either utilizes an exhaustive search through the metasurface parameter space or a priori intuitive knowledge. However, when complex designs with more dimensions (degrees of freedom) are required to attain highly constrained optical modes, complete exploration of the design space becomes computationally implausible and unrealistic. Moreover, intuitive methods cannot guarantee a globally optimal solution. On the other hand, inverse design methods include topology optimization, machine learning techniques, and evolutionary algorithms [12]. As discussed in [12], topology optimization uses gradient-based local optimizers that are not guaranteed to find the globally best solution since the metasurface parameter space typically involves several local optima. In machine learning techniques, a large dataset must be generated using electromagnetic simulations. The required training data for a problem scales up exponentially as device complexity increases through dimensions [13].

Interestingly, evolutionary algorithms provide a reliable solution to inverse design through stochastic numerical optimization. Methods like genetic algorithms [14], Ant-Colony Optimization (ACO) [15], and Particle Swarm Optimization (PSO) [16] have been applied to metasurface design. In particular, the Covariance Matrix Adaptation – Evolutionary Strategy (CMA-ES), based on adaptive sampling, provides faster convergence, more accurate results, is easily amenable to constraints, and can efficiently handle high-dimensional, non-separable and ill-conditioned problems with complex, non-linear and noisy objective function [12, 17-20]. An application of CMA-ES to 3D GaN phase gradient metasurfaces made of nanopillars of different shapes, targeting maximum light deflection efficiency, is demonstrated in Elsawy *et al*. [17]. However, for PCS-based applications discussed earlier, a general design method needs to be developed. In this article, we develop such a method by combining CMA-ES and Rigorous Coupled Wave Analysis (RCWA) [21] – a highly efficient Maxwell's equation solver for layered structures invariant along the z direction. The proposed approach significantly improves the efficiency of photodetector design by eliminating the need for exhaustive searches in high-dimensional parameter spaces. The successful development of such a tool holds significant promise for various applications in photonics, communication, and sensing, where tailored light absorbers are in high demand. In what follows, we enunciate our method, approach, and results.

## 2. Inverse design approach

Our approach is to systematically optimize the PCS design for controlling the parameters absorbance, reflectance, and transmittance in the desired optical modes, either all of them simultaneously or otherwise. This method is agnostic to the choice of materials and the targeted optical modes characterized by wavelength, polarization, and angle of incidence. Although the methodology and strategy can be easily adapted to any of the optical parameters and properties of the systems, i.e., tailoring absorbance, reflectance, and transmittance, we limit this discussion to designing photodetectors with increased absorbance.

To demonstrate our algorithm, we implement a three-layered photodetector design made of 7-layered (4 nm) black Phosphorus (bP) [22] as a 2D material for light absorption, non-absorbing silicon (with real and imaginary parts of the refractive index $n = 3.5$ and $k = 0$, respectively) as the material for the photonic crystal substrate, and a metallic mirror reflector at the bottom, as shown in Fig. 1.

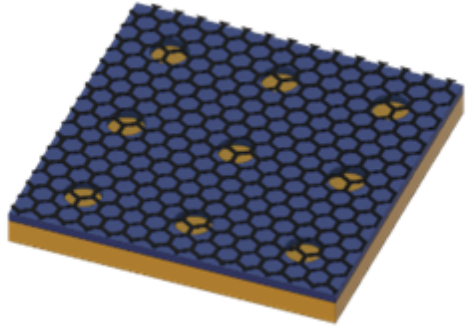

Fig. 1. This schematic represents the three-layered structure used to design photodetectors. The top mesh represents the 4 nm thick 7-layered black Phosphorus, the blue layer with holes is the patterned silicon metasurface, and the bottom yellow layer represents a lossless metallic mirror.

Given the following information,
   a) refractive indices of all materials denoted by $\Theta$,
   b) a choice of geometrical shapes and patterns,
   c) a set of feasible values for all dimensions of the metasurface denoted by $\Omega$, and
   d) a set of N target wavelengths and optical modes $\Lambda = \{\lambda_1, \lambda_2, \ldots, \lambda_N\}$

we need to design a metasurface that supports resonances, leading to high absorption at these wavelengths. In this case, we use a cylindrical hole with a circular cross-section in Si, i.e., a 2D periodic square array of unfilled air holes in Si, for designing the PCS. However, the same method can also be applied to any other cross-sectional pattern, like polygonal holes or apertures. We chose the circular cross-section with smooth edges since these are realistic to fabricate as against those with sharp corners/edges. Thus, the optimizable geometric parameters, as shown in Fig. 2, are periodicity $l$ of the square lattice, thickness $t$ of the PCS, the position of the centre of the hole in the unit cell $c = (c_x, c_y)$, and the radius of the corresponding hole r. Further in this study, we allow for the possibility of a number of holes $N_0 > 1$ in the unit cell, even allowing partial hole as in Fig. 2.

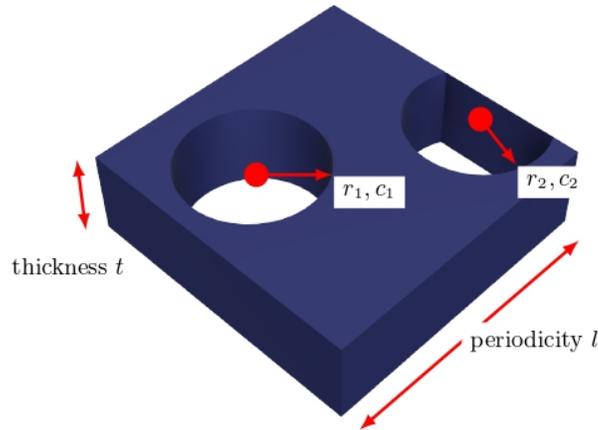

Fig. 2. A schematic showing the metasurface unit cell and its optimizable geometric parameters – thickness $t$, periodicity $l$, hole radii $(r_1, r_2)$, and hole centers $(c_1, c_2)$ – a case with more than one hole. In this case, the second

hole partially occupies the unit cell. Upon building a 2D periodic lattice of these unit cells, the desired PCS is created.

For optimization over multiple target wavelengths or optical modes, i.e., cases where N > 2, we allow for $N_0 > 1$ holes per unit cell to have more degrees of freedom and, thus, a higher dimensional search space to find the optimal structure. These holes may or may not be allowed to intersect. Therefore, in general, we will have $3N_0 + 2$ optimization variables - the position of the center within a unit cell and the radius, of each hole, as well as the periodicity and thickness of the PCS, each of which is bounded to remain within a range of feasible values, signifying box constraints, $\Omega_i$ for $i = 1,2,...,3N_0 + 2$. By specifying such a feasible set for each dimension, we can control and limit the search space. Let $X$ denote the vector of geometry variables. We have $X \in \Omega \subset \mathbb{R}^{3N_0+2}$. As mentioned, absorption would be a function of these geometric parameters and the material refractive indices. Formalizing this mathematically, we define the absorption function for the photonic crystal A: $(\lambda, \Omega, \Theta) \to [0, 1]$ for wavelength $\lambda \in \Lambda$. To find the best conditions for employing this PCS as the active component of a photodetector, we wish to find the optimal metasurface geometry $X^*$ such that,

$$X^* = \arg \max_{X \in \Omega} \sum_{i=1}^{N} A(\lambda_i, X, \Theta)$$

*2.1 Electromagnetic calculations*

To calculate the optical response of our metasurface designs, we employ rigorous coupled wave analysis (RCWA) [21]. This technique solves Maxwell's equation in the frequency domain by spatial discretization in the direction normal to the layer, whereas, within each layer, they are solved analytically in the Fourier domain. RCWA is a well-established method for analyzing layered periodic structures, favored for its high computational efficiency and speed. In this work, we performed all simulations using a Python implementation of RCWA known as GRWCA, cf. ref. [23]. Additionally, we leverage Poynting vectors calculated at layer boundaries to determine the layer-wise absorption profile within the metasurface.

*2.2 Optimization*

The absorption function A, as defined above, is a complex, non-convex, non-separable, and rugged (multiple local optima) function, the direct gradient of which is not explicitly computable. Thus, we require a stochastic numerical optimization technique. As discussed in [17], to find global maxima in this case, the CMA-ES stochastic optimization algorithm proves to be a particularly well-suited tool due to the nature of the absorption function. For a detailed explanation of the algorithm, we refer the reader to references [24-26].

Essentially, the CMA-ES algorithm treats all the design parameters as random variables drawn from a multivariate Gaussian distribution. The core concept herein lies in iteratively evolving the mean and covariance matrix of this distribution to guide the search towards promising regions of the design space. During each iteration, CMA-ES updates the mean of the distribution to favor previously successful candidate solutions, i.e., those with better objective function values. Simultaneously, the covariance matrix is also adaptively updated to favor search directions that have led to past improvements. This algorithm explained in Fig. 3, requires an initial mean vector $m_0$ constituting initial values for all the parameters in the metasurface geometry $X$. Each metasurface parameter is subject to a bound constraint specified by $\Omega_i$ introduced earlier, defining a continuous range of allowed values. We initialize the unit cell dimensions, namely, its period and thickness, within this search space and the hole radii with random values. Notably, in the case of a single hole ($N_0 = 1$), the hole center is initially positioned at the center of the unit cell, corresponding to complete holes within each unit cell, which is the preferred solution. However, the optimization process can shift the hole center, potentially leading to partial holes, see Fig. 2, depending on the specific problem and the optimal design. Altogether, this creates the mean vector $m_0$, an initial guess composed of initial values for all design parameters.

In addition to $m_0$, the algorithm requires a hyperparameter $\sigma_0$ – the initial step-size. The CMA-ES algorithm dynamically adapts the step-size using cumulative step-size adaptation (CSA). This step-size controls the search "reach" in each iteration, balancing the exploration of new regions in the initial stages with focused exploitation around promising areas as the search progresses. $\sigma_0$ and $m_0$ should be such that the optimum presumably lies within the initial high-dimensional cube $m_0 + 3\sigma_0(1, ..., 1)^T$. It is suggested that given a search space $\Omega_i = [a, b]$, one may start with $\sigma_0 = 0.3(b - a)$ [24]. In our experiments, we typically used $0.5 < \sigma_0 < 1.0$. Overall, the hyperparameters and initializations required by the algorithm are $m_0, \sigma_0, \Omega_i$ for each variable, and the number of holes per unit cell $N_0$. However, an informative guess for $m_0$ would often lead to the best results faster than a random $m_0$, i.e., the convergence time depends on this starting point. For instance, if the initial hole radius is too small or too large for a small unit cell, finding the solution may be difficult as the PCS might not play its role. Due to the stochastic nature of CMA-ES and the random start, multiple optimization trials, typically 5-10, might be necessary to achieve the best design. Here, we note that due to the nature of the algorithm's code implementation in Python, the objective function is minimized by default. Hence, we minimize the sum of reflectivity and transmissivity to maximize absorptivity for a layer.

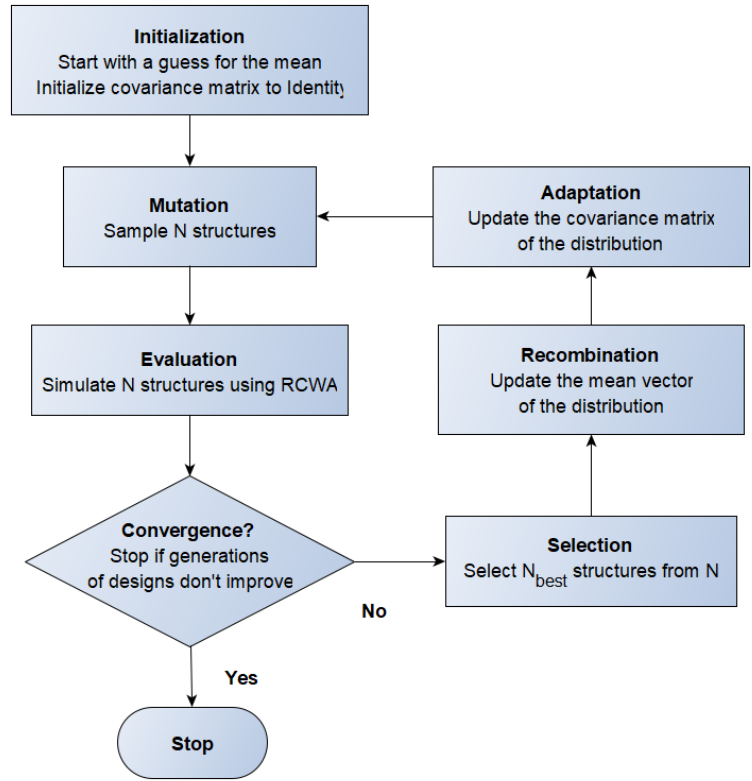

Fig. 3. The CMA-ES Algorithm for photodetector design as explained in [17]. We start with an initial guess for the mean of all design variables. From a few samples drawn from the Gaussian and simulated through RCWA, the best few are selected to update the mean and covariance. The distribution keeps updating until no significant improvement is observed in sampled designs.

The algorithm stops on a few termination criteria related to numerical stability described in [24]. The ones most relevant for this application are stagnation and function tolerance. The stagnation criterion checks if the designs have not improved over several generations by comparing the median of the last few values with the first. For a minimization problem, the tolerance criterion halts the process if the function value falls below a threshold, typically $10^{-10}$. Exploiting the scale invariance of Maxwell's equations, we normalize all lengths used within the optimization process to a base unit length of 1 micron. This allows us to treat them as dimensionless quantities that can be processed for numerical optimization. In the following section, we further describe present representative results that demonstrate power and application of this method.

## 3. Results and discussion

The following simulations were performed on a single core of the Dual Intel Xeon Gold 6248, 20-core 2.5 GHz CPU Node at the High-Performance Computing Facility, Indian Institute of Technology, Madras. Typically, the optimization runs may require from a few minutes to two hours of runtime, depending on the problem's complexity. The code we developed and used in this work is open to access, as we show in Code 1 (Ref. [32]).

*3.1 Single resonance at 1.55 μm*

Efficient light absorption at specific wavelengths in the infrared range is crucial for various applications in telecommunications and optoelectronics. The wavelengths of 1.55 μm and 2.1 μm hold particular importance for free space communication. Additionally, in optical communication systems, 1.55 μm coincides with the minimum attenuation window in silica fibers, making it the dominant wavelength for long-distance data transmission [27]. Efficient absorbers at 1.55 μm are essential for developing critical components like sensors, optical modulators, and terminators within these systems. Photodetectors operating at an atmospheric transmission wavelength of 2.1 μm are crucial for applications such as imaging [28], sensing [29], and high-speed communication [30].

The bP exhibits a highly tunable bandgap, ranging from 0.3 to 2 eV [31]. This tunability allows for the choice of the material for the bandgap required for a chosen wavelength. bP is chosen to be an active absorber for infrared photodetectors considered here. In our studies, we use seven layers (4 nm thick) of bP and the measured extinction coefficient [22]. We assume a normally incident plane wave on the absorber and optimize the absorption at the target wavelengths. Note that bP exhibits material dispersion, i.e., wavelength-dependent refractive index in this band. The proposed method can easily and efficiently accommodate material dispersion using a look-up table of finely sampled wavelengths and corresponding indices. However, we used a fixed value of index for silicon as the index changes very little in the bands considered in this study.

Our design for high absorption in 1.55 μm, schematically shown in Fig. 1, consists of bP on Si PCS with a metallic reflector. We used a 2D periodic array of one hole per unit cell ($N_0 = 1$) and optimized the parameters periodicity, thickness, hole radius, and hole center for the highest absorption. Fig. 4(a) shows nine-unit cells of the optimized metasurface design obtained from CMA-ES in 7 minutes of runtime on default settings (stop if change in absorption is less than $10^{-11}$). The absorption spectrum, calculated by integrating the Poynting vector in bP, is plotted as a function of wavelength in Fig. 4(b) for different number of iterations during the optimization process. Even starting with a random initial guess ($l = 728$ nm, $r = 408$ nm, $t = 148$ nm) with a poor absorption spectrum (black solid curve), the proposed method converges to an optimized structure (Fig. 4(a)) with the desired response ($l = 823, r = 253, t = 131$). We achieve 95% absorption in about a minute after 24 iterations (dotted blue curve) and 100% absorption after 228 iterations (solid red curve). The optimized structure shows a broad full width at half maximum (FWHM) absorption bandwidth of 146 nm. When the optimized design parameters are rounded off to the nearest 5 nm, the calculated absorption (dotted red curve) changes only very little, indicating that the design is robust to fabrication variations. Additionally, we noted that this design with such a large absorption bandwidth is robust also against temperature variations. For example, the refractive index of silicon reduces by 0.05 to 3.45 at 50K [33] and correspondingly, the peak redshifts by 15 nm, which is well within the bandwidth.

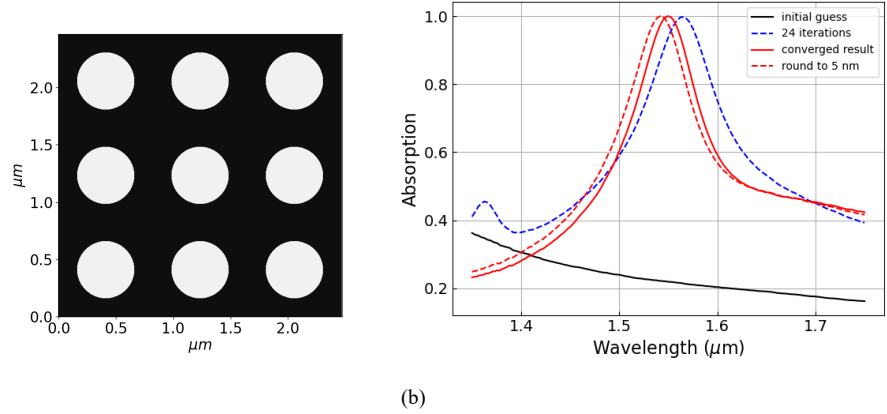

(a)            (b)

Fig. 4. (a) The periodic array ($r = 253$ nm, $l = 823$ nm) of holes in silicon ($t = 131$ nm) for a photodetector operating at 1550 nm. (b) Absorption in bP. The black curve indicates the absorption spectrum for our initial random guess. By the 24$^{th}$ iteration, we achieve >95% absorption at 1550 nm (blue dotted), in just about a minute of simulation time. The solid red curve shows the final result after 228 iterations corresponding to the metasurface in (a). The re-calculated absorption spectrum with rounded off the values for $r$, $l$, and $t$ to the nearest 5 nm is also shown (red dotted).

## 3.2 Single resonance at 2.1 μm

To design a photodetector for 2.1 μm, we replaced the metallic reflector with a distributed Bragg Reflector (DBR) mirror made of alternating SiO$_2$ and Sb$_2$S$_3$ layers (Fig. 5). Since a DBR is not reflective to all wavelengths, this design is aimed to test if the algorithm remained effective even without a perfect and broadband reflector. In this case too, the metasurface design variables $X$ remain the same as before. Figure 6(a) shows the optimized metasurface design generated in just 20 minutes of runtime. The calculated absorption in bP is plotted in Fig. 6(b). We see that absorption at the designed wavelength of 2.1 μm is 100% but with a narrower (±10 nm) band of absorption of over 80%. Additionally, we see unintended resonances at a few nearby wavelengths. Clearly, it is advantageous to have a broadband reflecting ground in place. Nevertheless, perfect absorption with a monolayer of material can be achieved leading photodetector with negligible dark current and near-perfect photocurrent.

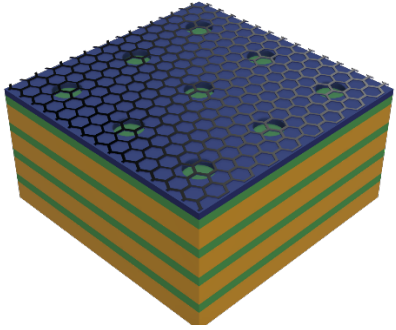

Fig. 5. Schematic of the system used to design a photodetector for 2.1 um. The top mesh represents black Phosphorus; the blue layer with holes is the patterned silicon metasurface. The metal layer from the previous design (Fig. 1) is replaced with a DBR mirror of alternating SiO$_2$ (green) and Sb$_2$S$_3$ (orange) layers.

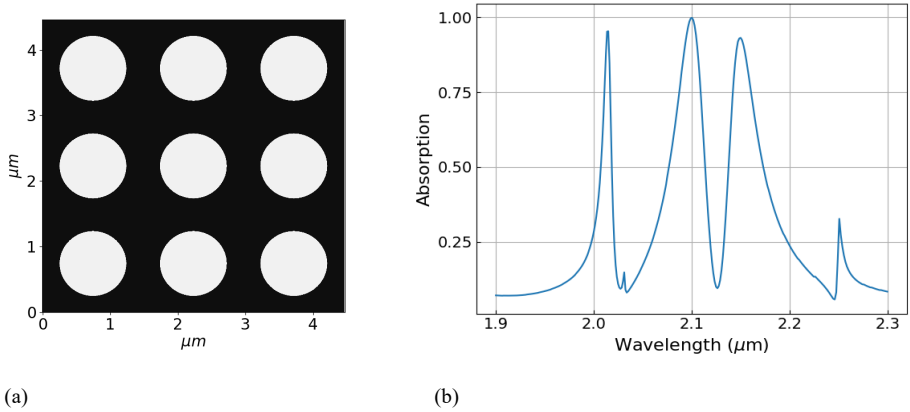

(a)            (b)

Fig. 6. (a) The periodic array ($r = 499$ nm, $l = 1.48$ μm) of holes in silicon ($t = 113$ nm) for a photodetector operating at 2.1 μm. (b) Absorption in bP.

*3.3 Resonance at two wavelengths*

Next, we apply our method to design a PCS that supports double (or multiple) resonances at the chosen wavelengths 1.3 μm and 1.55 μm. This design is highly desirable for seamless channel switching and simplified system design in optical communications [9]. Using the same three-layered structure (bP, silicon, metal as in Fig. 1), we employed the CMA-ES technique to design double resonance. We kept $N_0 = 1$ to enable faster search in a small search space. The nine unit cells of the metasurface obtained through 53 minutes of optimization runtime are shown in Fig. 7(a). Note that the unit cell does not contain one full hole. This creates a bit complicated structure for fabrication as a trade-off for a larger search space to solve a more demanding optimization problem. The absorption spectrum of this optimized structure is plotted in Fig. 7(b), which shows resonant absorption peaks in bP at both wavelengths of interest. The absorption is nearly 100% at 1.55 µm but is only ~80% at 1.3 µm. A larger search space (more complicated design) is required to get near-perfect absorption at both wavelengths.

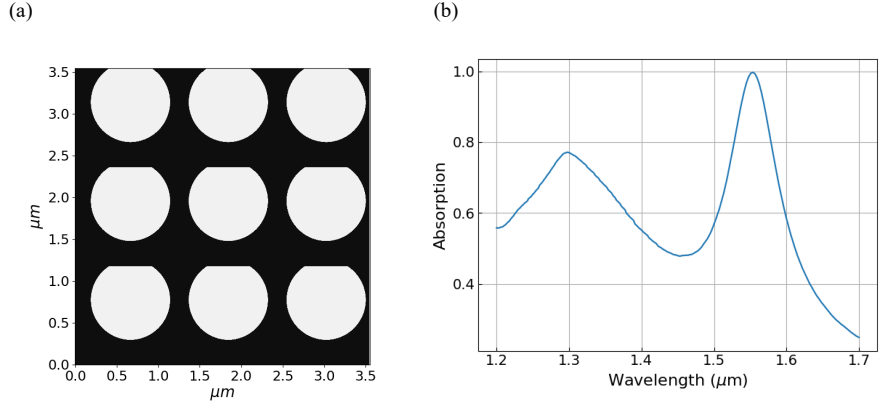

Fig. 7. (a) The silicon PCS design ($r = 482$ nm, $l = 1.18$ µm, and $t = 109$ nm) for high absorption at 1.3 and 1.55 µm. As the number of target absorption wavelengths increases, more degrees of freedom are required. We allow the hole to partially leave the unit cell to enable that, thus creating an asymmetric structure with partial holes. (b) Absorption in bP.

*3.4 Absorption at wide incident angles*

So far, we have optimized the metasurface for normal incident plane waves. However, often the detectors will be realistically required to have high absorption for an optical mode finite spot size comprised of plane waves with wave-vectors lying in a cone of incidence. Here, we show that the same inverse design optimization procedure can be used not only to design and improve a spectral response but also an angular response. Such a design for 1.55 µm is illustrated in Fig. 8. Using the three-layered structure as before, we designed a metasurface that will enable wide-incident angle absorption in bP (Fig. 8b). To do so, we uniformly sampled a few points in the range $[0, \theta_{max}]$ and try to maximize absorption at each of these discrete angle samples. One may use any type of incident light polarization – TE or TM, or a combination of both, in RCWA. However, in this example, we use s-polarized (TE) light. We observe that the designed structure exhibits at least 80% absorption in a 30-degree cone of incidence, whereas the FWHM exceeds 40°. This optimization was performed in 90 minutes of runtime on default settings (change tolerance = $10^{-11}$).

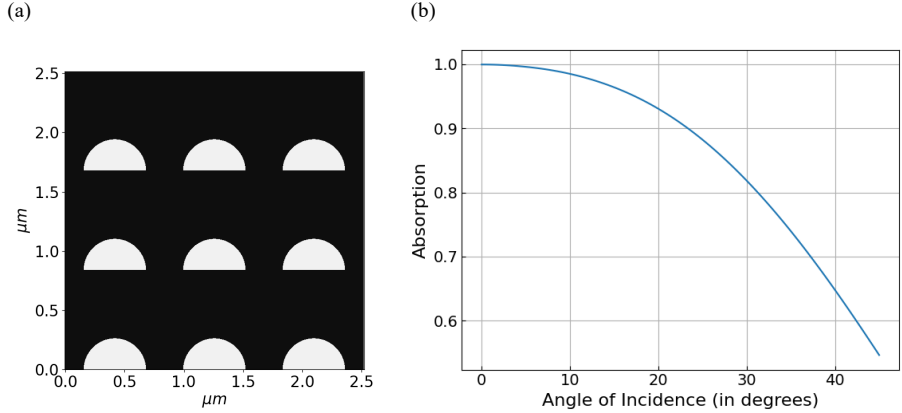

Fig. 8. (a) The periodic array ($r = 266$ nm, $l = 838$ nm) of almost semi-circular holes in silicon ($t = 126$ nm) for a photodetector absorbing a wide-incident angle cone of s-polarized light at 1.55 µm. (b) Absorption in bP plotted against the angle of light incidence.

As seen in the optimization for double resonant wavelengths and wide-angle absorption, we traded-off hole symmetry and introduced complexities in the design to meet more stringent requirements. When more resonances are desired, more complicated structures may be required. Furthermore, one may tune hyperparameters like $N_0$ to have two or more such holes per unit cell. One could also work with a triangular lattice of holes. However, like any practical optimization problem, there will be a practical limit to the number of different resonances that can be designed. This heavily depends on the choice of materials and pattern type from a physics perspective. Nevertheless, limitations can be tackled by experimenting with hyperparameters, patterns, and the objective function used to frame the optimization problem.

For a given absorption spectrum, if a larger bandwidth is needed, we can choose the two target wavelengths that are a bit closer to each other so that the width of each peak can overlap to form broader absorption. Eventually, the absorption bandwidth is limited by the absorption coefficient of the absorber and can be improved by suitable choice of absorber.

Multiple avenues exist to improve this inverse design procedure through mathematics and physics. One can experiment with various metasurface layers by optimizing them together or sequentially. Furthermore, a staircase approximation is required to simulate conical holes or curved cavities as RCWA only deals with layers invariant along the z-axis. However, this can quickly become expensive. Mathematically, in the absorption maximization problem (Section 2), we can explore possibilities by changing the summation to a weighted sum or a product of individual objective functions. Interestingly, it might also be possible to draw inferences from the photonic band diagrams of such structures to gain deeper insights into how the shape of the cavity influences optical response. This analysis might be instrumental when an excellent initial guess is required; the algorithm can further improve upon that.

## 4. Conclusion

In conclusion, we have developed a powerful technique for inverse designing periodic dielectric metasurfaces by efficiently optimizing the geometric patterns and dimensions of the metasurface. As a demonstration of the method, we applied it to design metasurface structures to achieve near 100% absorption at the chosen wavelengths of 1.55 µm and 2.1 µm with a monolayer of 2D material. The ability to achieve near perfect absorption at single wavelength will be useful in designing single photon avalanche photo detectors operating at room temperature. In addition, this method is applied to achieve double resonances in an integrated cavity structure to demonstrate its possible use in efficient pump-probe, second harmonic conversion, and biphoton production experiments.

**Acknowledgments.** A. M. J. acknowledges support from the Young Research Fellowship Program at IIT Madras. S. R. K. gratefully acknowledges support from financial support of the Dept. of Science and Technology (DST), Govt. of India (GoI), through the Core Research Grant of the Science and Engineering Research Board, Indo-German DST-DAAD bilateral scheme together with the German Academic Exchange Service (DAAD), and the Advanced Manufacturing Technology scheme of the Technology Development Board; he gratefully acknowledges support from the Ministry of Education (GoI) through the Scheme for Transformational and Advanced Research in Sciences (STARS) project grant and through the Institute of Eminence (IoE) scheme (project SB20210813PHMHRD002720), which funds the Quantum Center of Excellence for Diamond and Emergent Materials (QuCenDiEM). Rituraj acknowledges support from Science & Engineering Research Board through project number SRG/2022/001090. SK thanks IIT Madras for support through a visiting position. We also acknowledge the use of the computing resources at HPCE, IIT Madras.

**Data availability.** Data underlying the results, and the code used for simulations presented in this paper are available at Code 1 (Ref. [32]).

**Disclosures.** The authors declare no conflicts of interest.